\documentclass[reprint,aps,pra,twocolumn,showpacs]{revtex4-2}

\usepackage{graphicx}
\usepackage{dcolumn}
\usepackage{bm}
\usepackage{hyperref}
\usepackage[mathlines]{lineno}

\hypersetup{colorlinks=true, citecolor=blue, urlcolor=blue, linkcolor=blue}
\usepackage{slashed}  

\usepackage{amsmath}
\usepackage{amssymb} 
\usepackage{natbib}
\usepackage{mathrsfs}
\usepackage{flushend}
\makeatletter

\newcommand{\Rmnum}[1]{\expandafter\@slowromancap\romannumeral #1@}
\makeatother
\allowdisplaybreaks[3]

\begin{document}
\preprint{APS/123-QED}

\title{Integrating the advantages of two single-pixel imaging schemes via holographic projection in ghost-imaging systems}



\author{Liming Li}
\email{liliming@sdut.edu.cn}
\author{Zhenguo Zhao}
\author{Gongxiang Wei}
\author{Wenfei Zhang}
\author{Huiqiang Liu}
\email{liuhq@sdut.edu.cn}

\affiliation{School of Physics and Optoelectronic Engineering, Shandong University of Technology, Zibo 255049, China\\}


\date{\today}

\begin{abstract}
Computer-generated hologram (CGH) allows for the on-demand scaling and projection of artificially designed target patterns, while incorporating benefits such as a lensless setup and high-frame-rate operation. In this work, we actively control the projection pattern using CGH and integrate two typical single-pixel imaging (SPI) schemes, thereby implementing a ghost imaging (GI) scheme with flexibly tunable properties. Specifically, various reference signals from computational holography and the corresponding bucket signals are used in the intensity correlation algorithm. Accordingly, those GI results enable the parallel presentation of the outcomes from these two SPI schemes. In the experiment, two types of target patterns, intensity-squared chaotic speckle and artificially designed sparse matrix, are used to perform GI. Those imaging results indicate a significant improvement in ghost image visibility, irrespective of whether the reference signal is the reconstruction or target pattern of computational holography. Furthermore, we realize positive and negative copies of ghost image via holographic projection in which symmetrical mirror target patterns are artificially designed. Thus, by integrating these two SPI schemes, the lensless GI scheme based on CGH not only advances towards the visibility requirements for practical applications but also enables a high-frame-rate projection scheme essential for multi-frame intensity correlation measurements.
\end{abstract}
\maketitle

\section{Introduction}
In 1995, Pittman \emph{et al}. have realized the second-order ghost imaging (GI) using a quantum light source~\cite{Pittman1995Optical}. After nearly a decade of research~\cite{Bennink2002Two,Bennink2004Quantum,Gatti2004Correlated,Cheng2004Incoherent,Ferri2005High,Valencia2005Two,Zhang2005Correlated,Cao2005Geometrical}, it has been demonstrated that quantum source is not a necessary condition but classical source, such as a pseudo-thermal light generated by laser beam passing through a rotating ground glass~\cite{Ferri2005High,Valencia2005Two}, is a cost-effective alternative for GI. Traditional ghost imaging (TGI) with pseudo-thermal light requires two dynamic optical signals, reference signal (RS) and test signal (TS), for intensity correlation~\cite{Bennink2002Two,Bennink2004Quantum,Gatti2004Correlated,Valencia2005Two,Zhang2005Correlated,Scarcelli2006Phase,Basano2006Experiment,Cai2007Lensless,ChenOL2010Highvisibility,Li2019Transverse}. After passing through a beam splitter, one beam undergoes free diffraction and creates a dynamic speckle acting as the RS, while the other beam transmits through the detected object to form the TS, which can be measured by a single-pixel detector. Due to the disordered transmissive surface of ground glass~\cite{Martienssen1964Coherence,Goodman2007Speckle}, the RS can only be obtained by experimental measurement. Thus, TGI with pseudo-thermal light requires two dynamic intensity signals: the TS and the RS, both of which are indispensable.

After introducing a liquid-crystal spatial light modulator (SLM) into the experimental setup, the RS can be theoretically calculated using the Huygens-Fresnel (HF) principle~\cite{Goodman1995Introduction}, leading to a new scheme named computational ghost imaging (CGI) (or single-pixel imaging (SPI))~\cite{Shapiro2008Computational,BrombergPRA2009Ghost,SunScience20133D,Klein2019Beam}. The CGI and relevant improvement schemes~\cite{Katz2009Compressive,Clemente2010Optical,Erkmen2012Computational,Katkovnik2012Compressive,Abetamann2013Compressive,Devaux2016Computational,Shimobaba2018Computational,Wu2020Deep,Sui2021Optical,Wu2020Online,Zou2023Target,Rosi2024Increasing,Wang2024High,Lin2025Demonstration} not only simplify the experimental setup to establish the SLM-SPI scheme but also incorporate information theory into the imaging algorithm, leading to a reduction in the sampling rate without sacrificing image quality, such as compressed sampling~\cite{Katz2009Compressive,Katkovnik2012Compressive,Abetamann2013Compressive,Sui2021Optical} and deep learning~\cite{Shimobaba2018Computational,Wu2020Deep,Zou2023Target}, etc. 

The modulation rate of a liquid-crystal SLM is typically 60 Hz. By contrast, a commercial digital micromirror device (DMD, DLP Discovery D4100 kit, Texas Instruments, USA) can operate at rates as high as 32 kHz for binary amplitude-only CGH. Inspired by SLM-SPI scheme, an alternative approach of DMD-SPI scheme has also emerged that replaces conventional thermal speckle with artificially designed structured light patterns~\cite{Duarte2008Single,Shrekenhamer2013OETerahertz,HornettNL2016Subwavelength,Edgar2019Principles,GibsonOE2020Single-pixel,Xiao2022OESingle,HuOE2026Fast}. Those patterns are generated by the programmable, high-frame-rate DMD array, then scaled and projected onto the detected object by using an optical lens. Although the measurement scheme is identical to that of SLM-SPI, the introduction of the optical lens in the DMD-SPI setup constrains its potential for practical application. Thus, a lensless SPI scheme with the features of arbitrary projection pattern design and a high-frame-rate operation is valuable for practical applications.

Here, we propose a holographic projection GI scheme that employs CGH with artificially designed target patterns instead of thermal speckles, enabling the parallel presentation of imaging results of those two SPI schemes (SLM-SPI and DMD-SPI). Although this work uses phase-only CGH, the scheme could be extended to grayscale or binary amplitude-only CGH using standard methods (e.g., Gerchberg–Saxton (GS) algorithm~\cite{Gerchberg1976A} and Otsu’s thresholding~\cite{OtsuAuto1975AThreshold}), though such extensions remain to be implemented experimentally. Therefore, the development of GI scheme via holographic projection holds great potential for practical applications.

In this article, we will first conduct a detailed comparison of correlated signals in the intensity correlation algorithm between thermal light GI and holographic projection GI we proposed. Then, we show the GI experiment by holographic projection in which target patterns originate from intensity-squared chaotic speckle and artificially designed sparse matrix. Those GI results demonstrate a substantial increase in the visibility. Finally, we achieved both positive and negative copies of the ghost image on the basis of the artificially designed symmetrical mirror speckle by holographic projection.

\begin{figure}[b]
	\centering
	\includegraphics[width=0.48\textwidth]{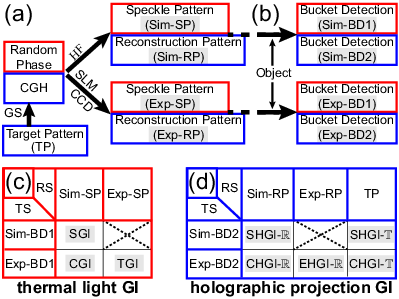}
	\caption{Generation processes of reference signals (a) and test signals (b) in thermal light GI (shown by red solid line boxes) and holographic projection GI (shown by blue solid line boxes) are represented by a flow diagram. TP: target pattern; GS: Gerchberg-Saxton algorithm, CGH: computer-generated hologram; HF: Huygens-Fresnel principle; SLM: spatial light modulator; CCD: charge coupled device. (c) and (d) are the correspondence relationship between the abbreviations of GI scheme and data sources of intensity correlation in thermal light GI and holographic projection GI, respectively. TS: test signal; RS: reference signal. See text for explanation of abbreviations on various reference signals, test signals and GI schemes. Here, key abbreviations in each subplot are highlighted with a gray background.\label{01GIName}}
\end{figure}

\section{Theoretical model} 
Figure~\ref{01GIName} presents the abbreviations for various thermal light GI schemes and holographic projection GI schemes, along with the correlated sources of their corresponding RS and TS. The sources of RS and TS are illustrated in Figs.~\ref{01GIName}(a) and \ref{01GIName}(b), respectively. In the thermal light case, the speckle pattern originates from the free diffraction of a random phase of thermal source. What's more, according to the GS algorithm, the reconstruction pattern detected by CCD camera in the holographic projection case can be realized by the free-diffraction of CGH loaded by SLM for artificially designed target pattern (TP)~\cite{Raymond2024Holography}. Here, Sim-SP and Sim-RP in Fig.~\ref{01GIName}(a) are the speckle pattern and the reconstruction pattern, respectively, by numerical simulation using the HF principle. Besides, Exp-SP and Exp-RP in Fig.~\ref{01GIName}(a) are experimental signals, corresponding to simulated signals Sim-SP and Sim-RP, respectively. What's more, Sim-BD1, Sim-BD2, Exp-BD1 and Exp-BD2 in Fig.~\ref{01GIName}(b) are TS after detected object inserted, corresponding to those RS, Sim-SP, Sim-RP, Exp-SP and Exp-RP, respectively.

Figures~\ref{01GIName}(c) and \ref{01GIName}(d) show the abbreviations in thermal light GI and holographic projection GI, respectively. In addition to TGI and CGI schemes the most common categories of thermal light GI, simulated ghost imaging (SGI) in Fig.~\ref{01GIName}(c), originated from the intensity correlation between the Sim-SP and Sim-BD1, is one of the GI schemes for judging imaging performance by numerical method~\cite{Basano2007Aconceptual,Sun2012Normalized,Wu2024High}. Due to the absence of noise influence, SGI is the optimal theoretical result among the three schemes. By comparing the data sources and the intensity correlation algorithm, the experimental holographic projection GI, the computational holographic projection GI and the simulated holographic projection GI can be abbreviated as EHGI-$\mathbb{R}$, CHGI-$\mathbb{R}$ and SHGI-$\mathbb{R}$, respectively. Here, the character '$\mathbb{R}$' represents RS coming from the \emph{Reconstruction} pattern. What's more, two novel intensity correlation schemes emerge in the holographic projection GI by replacing the reconstruction pattern with the target pattern. Thus, computational and simulated holographic projection GIs with RS coming from the \emph{Target} pattern can be abbreviated as CHGI-$\mathbb{T}$ and SHGI-$\mathbb{T}$, respectively. Note that, CHGI-$\mathbb{R}$ and CHGI-$\mathbb{T}$ directly correspond to two typical SPI schemes (SLM-SPI and DMD-SPI), respectively.

In theory, the ghost image can be expressed by the normalized second-order intensity correlation function as follows~\cite{ChenOL2010Highvisibility,Li2019Transverse,Katz2009Compressive}:
\begin{equation}\label{EQ01}
	g^{( 2 )}(x) =\frac{{\left\langle {\mathcal{RS}\left( x \right)\times \mathcal{TS}} \right\rangle }}{{\left\langle {\mathcal{RS}\left( x \right)} \right\rangle }{\left\langle \mathcal{TS} \right\rangle }},
\end{equation}
where $\mathcal{RS}\left( x \right)$ is the RS at position $x$ in the reference plane, $\mathcal{TS}$ is the TS collecting all the projected light intensity, and $\left \langle \cdots  \right \rangle $ is the ensemble average, respectively. For the evaluation of imaging visibility, the following formula can be employed:
\begin{equation}\label{EQ02}
	V=\left( g^{(2)}_{\text{max}}-g^{(2)}_{\text{min}} \right)/\left( g^{(2)}_{\text{max}}+g^{(2)}_{\text{min}} \right),
\end{equation}
where $g^{(2)}_{\text{max}}$ and $g^{(2)}_{\text{min}}$ are the maximum and minimum values of $g^{( 2 )}(x)$, respectively. Due to the noise present in the imaging results of GI, the extreme value is replaced by the mean of the adjacent measured points of ghost image in the experimental verification.

\section{Experimental verification} 
\begin{figure}[!t]
	\centering
	\includegraphics[width=0.48\textwidth]{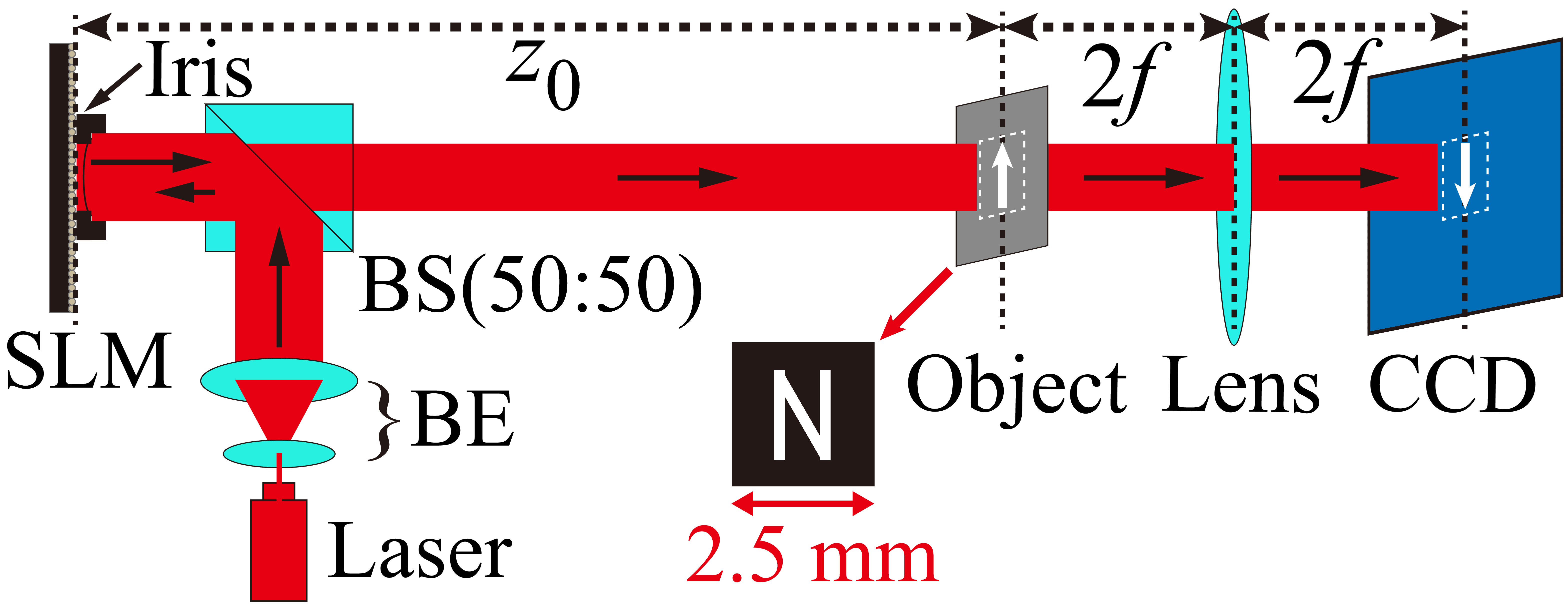}
	\caption{Schematic diagram of the experimental setup for the GI scheme. BE: beam expander; BS: 50:50 beam splitter; SLM: phase-only spatial light modulator; CCD: charge coupled device. $f$ is the focal length of the Lens. A letter "N" was served as an input object, placing at the central of the object plane of the $2f-2f$ imaging system.\label{02GIScheme}}
\end{figure}
Figure~\ref{02GIScheme} shows the experimental setup for the holographic projection GI with a phase-only spatial light modulator (SLM, an element pixel size of $12.3\times 12.3$  $\mu \text{m}^2$ and a total pixels $1280\times 1027$, GCI-770401, Daheng Optics, China). A single-mode 632.8 nm continuous-wave laser beam is expanded, collimated, and reshaped through a beam expander (BE). Then, the beam is reflected by a 50:50 non-polarized beam splitter (BS) and transmitted normally upon the effect window of SLM. The reflected phase-encoded laser beam successively transmits through the BS and arrives the detected object after being free diffracted over a distance $z_0=$ 54.6 cm between the SLM and the object. Here, a $2f-2f$ imaging system, created by the detected object, a lens, and CCD camera, performs the data measurement of those RS and TS described in the theoretical model, where $f=5.1$ cm is the focal length of the lens. We placed a 0-1 amplitude transmitting letter "N" with a slit width of 200 $\mu$m served as an detected object, placing it in the central of the object plane of the $2f-2f$ imaging system. As we can see, the TS can be acquired by summing the intensity values of those CCD pixels in the white dashed rectangle with the object in place, i.e. the total intensity transmitted through the object. In addition, the RS for experimental categories on the object plane is achieved by the CCD camera as the object is moved away. What's more, the effective diffraction size of the phase window in SLM is controlled by an iris with a diameter $D=4$ mm, which is placed as close as possible to the SLM. Given that the light field on the object plane originates from far-field diffraction of the phase-encoded laser beam in our work, a lens phase factor $\text{exp}\{-ik\vec{r}_{\text{SLM}}^2/(2z_0)\}$ was superposed with the input diffraction phase of SLM, where $k$ is the wave vector of light source and $\vec{r}_{\text{SLM}}$ is the spatial position of each pixel in SLM. For a simplified description, the superposition of lens phase factors will not be described in the text again.

\begin{figure}[!t]
	\centering
	\includegraphics[width=0.48\textwidth]{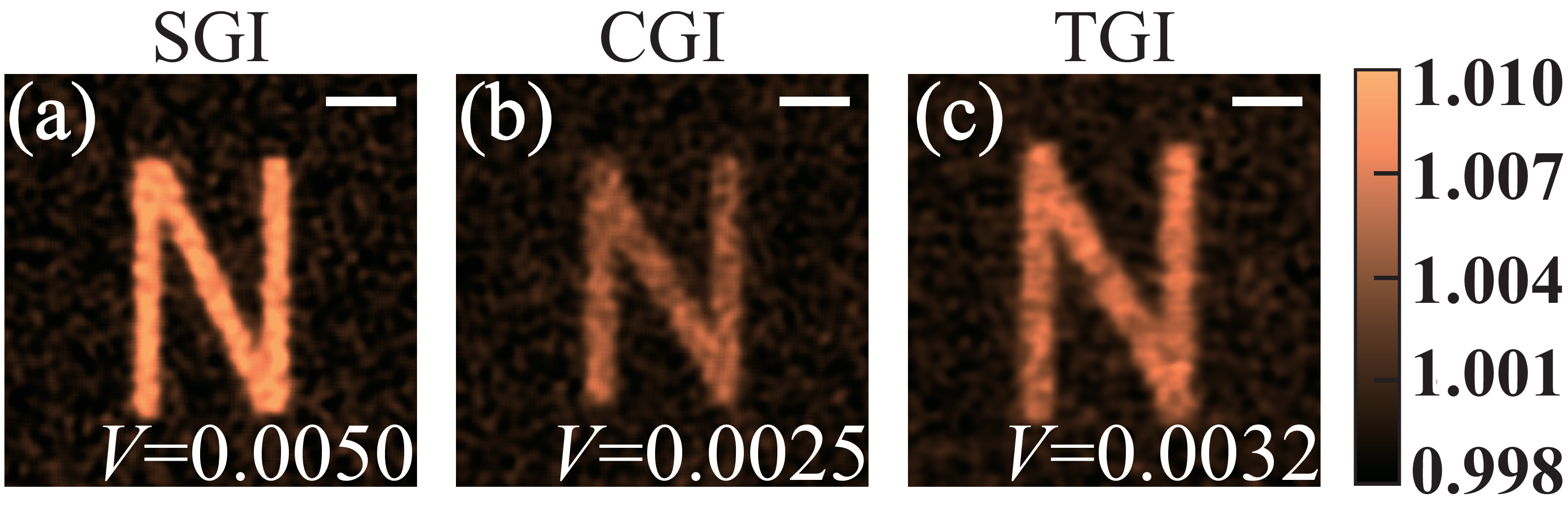}
	\caption{Simulated or experimental results of ghost image with chaotic source according to three GI schemes: (a) SGI, (b) CGI and (c) TGI, respectively. The visibility $V$ of ghost image are displayed in the bottom right corner of the subgraphs. Three subgraphs share a common color bar. The scale bars in all subgraphs are 400 $\mu$m. \label{03ThermalCase}}
\end{figure}

To compare with subsequent imaging results, we first utilized the same experimental setup to obtain the results of traditional thermal light GI. Given the single-mode of laser beam, we are limited to producing a chaotic source. To create the chaotic source by the phase-only SLM, a dynamic random phase $\text{exp}\{i \phi(\vec{r}_{\text{SLM}}, t)\}$ puts into the SLM and can produce a speckle sequence by far-field diffraction, where $\phi(\vec{r}_{\text{SLM}}, t)\in [0, 2\pi)$ is a uniform distribution random phase and completely independent of spatiality and time~\cite{Shapiro2008Computational,Li2019Transverse}. Three results of simulated or experimental reconstruction, SGI, CGI, and TGI, after 10,000 iterations are shown in Figs.~\ref{03ThermalCase}(a)-\ref{03ThermalCase}(c), respectively. Note that, three subgraphs in Fig.~\ref{03ThermalCase} share a common color bar. According to Eq.~(\ref{EQ02}), the visibility $V$ of SGI, CGI, and TGI in Fig.~\ref{03ThermalCase} are 0.0050, 0.0025, and 0.0032, respectively (displayed in the bottom right corner of the subgraphs). Compared to TGI, the CGI exhibits a slight reduction in visibility. Unlike TGI, the two-arm signals in CGI may not be perfectly identical due to the deviation of phase modulation of SLM. What's more, the noiseless SGI demonstrates the best imaging visibility, as our expected.

\begin{figure*}[htpb]
	\centering
	\includegraphics[width=0.65\textwidth]{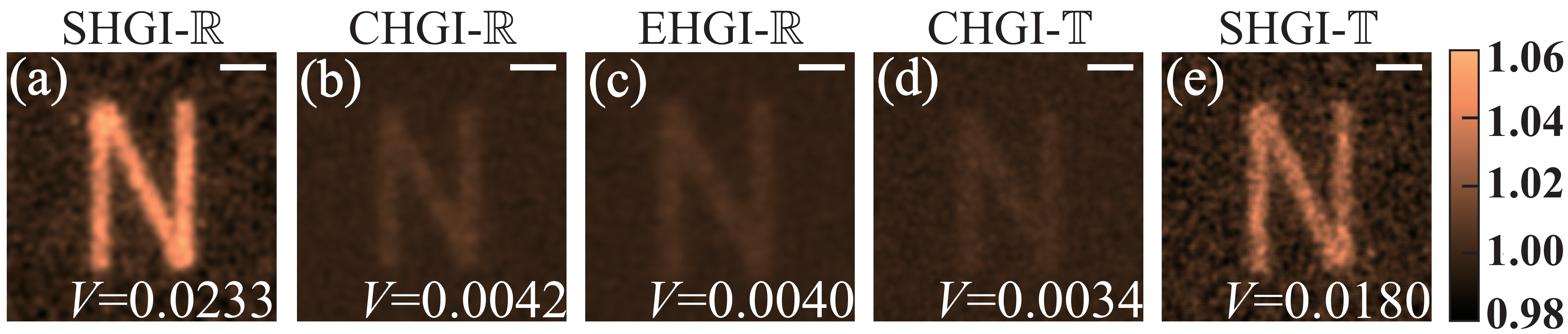}
	\caption{Simulated or experimental results of ghost image by holographic projection with the TP originated from the intensity squared-speckle according to five GI schemes: (a) SHGI-$\mathbb{R}$, (b) CHGI-$\mathbb{R}$, (c) EHGI-$\mathbb{R}$, (d) CHGI-$\mathbb{T}$ and (e) SHGI-$\mathbb{T}$, respectively. The visibility $V$ of ghost image are displayed in the bottom right corner of the subgraphs. Five subgraphs share a common color bar. The scale bars in all subgraphs are 400 $\mu$m. \label{04ThermalHoloCase}}
\end{figure*}

\begin{figure*}[htpb]
	\centering
	\includegraphics[width=0.75\textwidth]{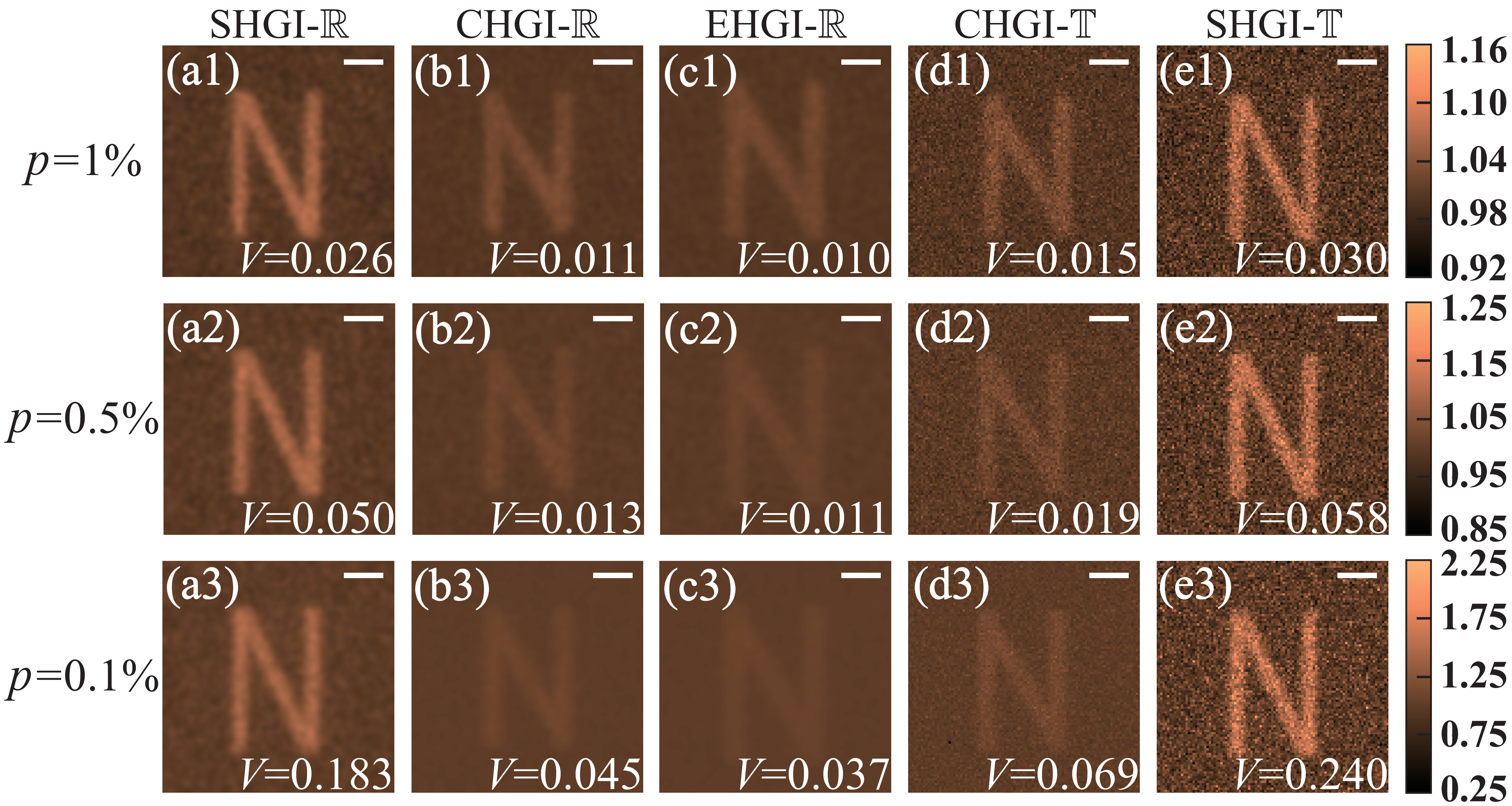}
	\caption{Five kinds of simulated or experimental holographic projection GIs with different sparsity parameter $p$. The visibility $V$ of ghost image are displayed in the bottom right corner of the subgraphs. To effectively display the imaging results, five subgraphs horizontally arranged with same parameter $p$ share a common color bar. The scale bars in all subgraphs are 400 $\mu$m.\label{05SparseHoloCase}}
\end{figure*}

In recent research~\cite{Li2025Active}, we have regulated the peak value of the Hanbury Brown-Twiss (HBT) effect based on phase-only CGH. Notably, when those TPs originate from intensity-squared chaotic speckle and artificially designed sparse matrix, corresponding Exp-RPs exhibit the super-bunching effect. As is well known, the HBT effect acts as the point spread function (PSF)~\cite{Shih2020AnIntroduction,Wu2024High,Liu2024Progress} of the GI scheme so that the super-bunching effect can enhance the visibility~\cite{ChenOL2010Highvisibility,Chan2009High,Zhou2019Experimental,Zhang2019Superbunching,Liu2021Simple,Ye2022Antibunching}. Building upon these results, we present the holographic projection GI. Here, each RS/TS is derived from the single-frame reconstruction patten of the corresponding TP in the holographic projection GI scheme. Details about the creation of those TPs and the second-order spatial correlation function of reconstruction patterns can refer to the Ref.~\cite{Li2025Active}.

Figure~\ref{04ThermalHoloCase} shows the holographic projection GI in the case of the intensity-squared speckle. Five results of simulated or experimental reconstruction, SHGI-$\mathbb{R}$, CHGI-$\mathbb{R}$, EHGI-$\mathbb{R}$, CHGI-$\mathbb{T}$, and SHGI-$\mathbb{T}$, after 10,000 iterations are shown in Figs.~\ref{04ThermalHoloCase}(a)-\ref{04ThermalHoloCase}(e), respectively. Here, five subgraphs share a common color bar. According to Eq.~(\ref{EQ02}), the visibility $V$ of SHGI-$\mathbb{R}$, CHGI-$\mathbb{R}$, EHGI-$\mathbb{R}$, CHGI-$\mathbb{T}$, and SHGI-$\mathbb{T}$ are 0.0233, 0.0042, 0.0040, 0.0034, and 0.0180, respectively (displayed in the bottom right corner of the subgraphs). Note that, the visibility of those simulated holographic projection GIs (SHGI-$\mathbb{R}$ and SHGI-$\mathbb{T}$) is superior to others. Although the visibility of those experimental results by CHGI-$\mathbb{R}$, EHGI-$\mathbb{R}$ and CHGI-$\mathbb{T}$ in Figs.~\ref{04ThermalHoloCase}(b)-\ref{04ThermalHoloCase}(d) using experimental bucket signals Exp-BD2 is relatively low, but still represents a significant improvement compared to experimental results of chaotic GI in Figs.~\ref{03ThermalCase}(b) and \ref{03ThermalCase}(c) by CGI and TGI schemes, respectively. Here, the advantage of visibility is not obvious, mainly because the experimental measurement of the high dynamic range of intensity-squared speckle is greatly affected by the environment noise. However, these simulated results in Figs.~\ref{04ThermalHoloCase}(a) and \ref{04ThermalHoloCase}(e) by SHGI-$\mathbb{R}$ and SHGI-$\mathbb{T}$ schemes avoiding the pollution of experimental noise demonstrate the potential of holographic projection-based GI for high-visibility.

\begin{figure*}[!htpb]
	\centering
	\includegraphics[width=0.65\textwidth]{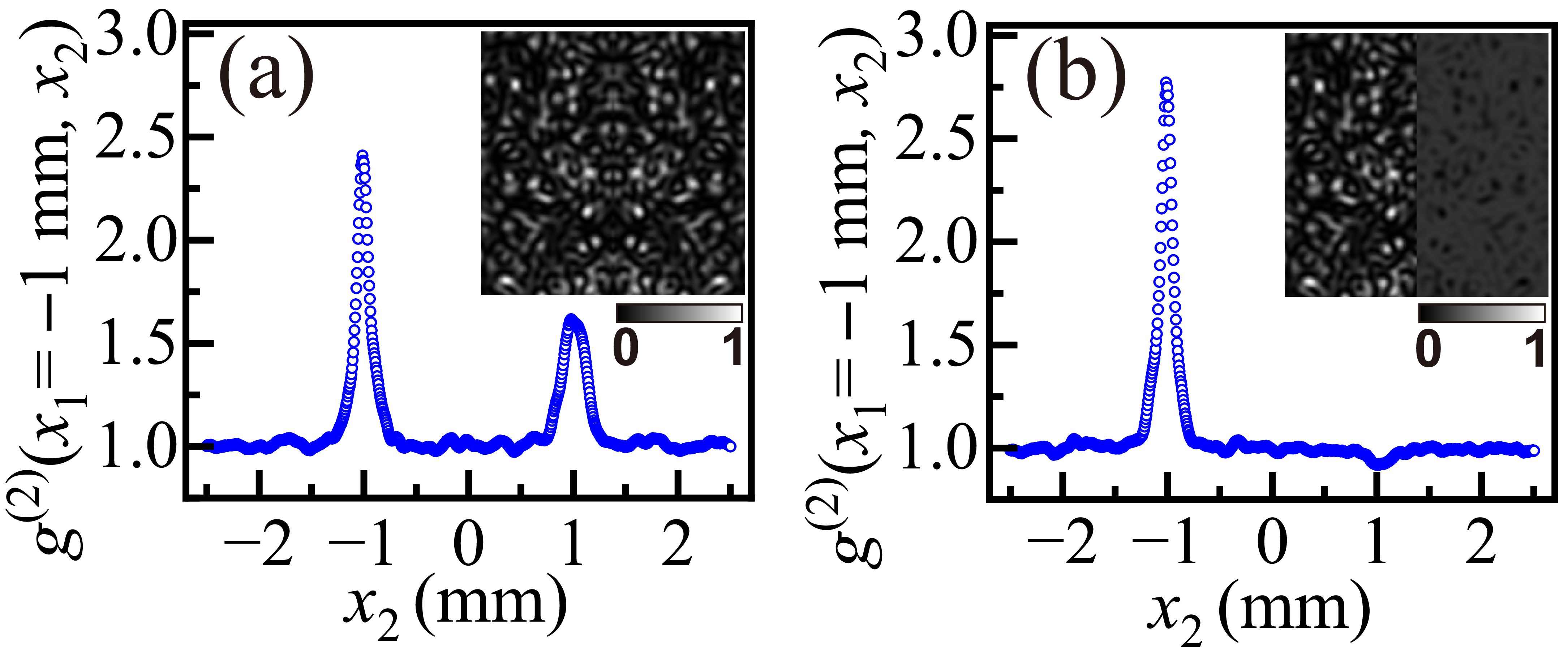}
	\caption{The experimentally measured HBT bunching curves of holographic reconstruction pattern after 10,000 iterations with positive (a) and negative (b) symmetric mirrors of normalized intensity-squared chaotic speckle. The inset at the upper right corner of each subgraphs is a single-frame TP. The inset includes a color bar.\label{06CopyCase}}
\end{figure*}

\begin{figure*}[htpb]
	\centering
	\includegraphics[width=0.70\textwidth]{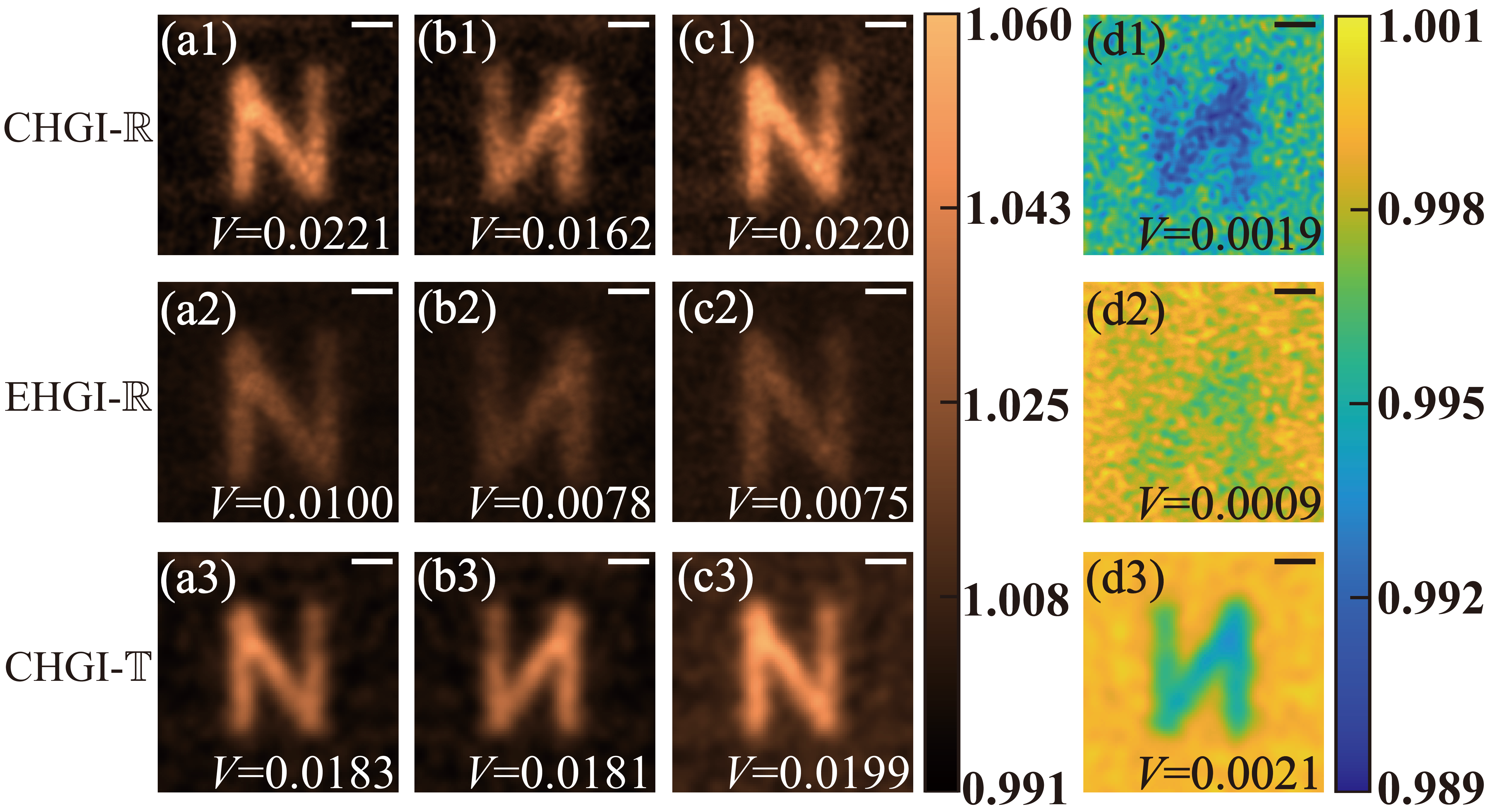}
	\caption{Three kinds of experimental holographic projection GIs with different TS/RS signals. (a1-a3) positive mirror self-correlated GI; (b1-b3) positive mirror cross-correlated GI; (c1-c3) negative mirror self-correlated GI; (d1-d3) negative mirror cross-correlated GI; The visibility $V$ of ghost image are displayed in the bottom right corner of the subgraphs. To effectively display the imaging results, two color bars are employed for display. The scale bars in all subgraphs are 400 $\mu$m.\label{07CopyCaseImage}}
\end{figure*}

To further enhance the ghost image visibility by holographic projection, we introduce three kinds of dynamic 0-1 binary sparse TPs, which are originated from an all-zeros matrix with a proportion $p=$ 1$\%$, 0.5$\%$ and 0.1$\%$, respectively, of 1 randomly inserted. As sparsity increases, the peak value $g^{(2)}(0)$ of bunching effect originating from the single-frame holographic reconstruction pattern increases quickly~\cite{Li2025Active}. Figure~\ref{05SparseHoloCase} shows five kinds of simulated or experimental holographic projection GIs with different parameter $p$ after 10,000 iterations. To effectively display the imaging results, five subgraphs with same parameter $p$ horizontally arranged share a common color bar. According to Eq.~(\ref{EQ02}), the visibility $V$ of those 15 subgraphs is displayed in the bottom right corner of the subgraphs. The specific numerical values are not repeated described in the text here. In general, as the parameter $p$ decreases (or sparsity increases), the visibility of holographic projection ghost images gradually improves. Similarly, the visibility of the simulated holographic projection GIs (SHGI-$\mathbb{R}$ and SHGI-$\mathbb{T}$) remains superior to that of the experimental types (CHGI-$\mathbb{R}$, EHGI-$\mathbb{R}$ and CHGI-$\mathbb{T}$) with the same parameter $p$. What's more, due to the superior holographic projection quality of sparse matrix than that of the thermal light speckle~\cite{Li2025Active}, CHGI-$\mathbb{R}$ and CHGI-$\mathbb{T}$ exhibit a better performance than EHGI-$\mathbb{R}$ in visibility. Therefore, the flexible design of the TP can further enhance visibility of image results in holographic projection GI. Overall, compared with those imaging results in the intensity-squared speckle case, holographic projection-based GI with TP originated from the sparse matrix provides better visibility. Especially, the visibility of each result with the parameter $p$ = 0.1$\%$ in Figs.~\ref{05SparseHoloCase}(a3)-\ref{05SparseHoloCase}(e3) exceeds the theoretical optimal result (by the SHGI-$\mathbb{R}$ scheme shown in Fig.~\ref{04ThermalHoloCase}(a)) of the intensity-squared speckle case.

A comparison of the GI performance of CHGI-$\mathbb{R}$ and CHGI-$\mathbb{T}$ schemes in Fig.~\ref{04ThermalHoloCase} and Fig.~\ref{05SparseHoloCase} reveals the fundamental difference for SLM-SPI and DMD-SPI: CHGI-$\mathbb{R}$ yields better results in Fig.~\ref{04ThermalHoloCase} with the intensity-squared chaotic speckle, whereas CHGI-$\mathbb{T}$ excels in Fig.~\ref{05SparseHoloCase} when the spares matrix is employed. The key distinction lies in the fundamental difference between the correlation signals on which those two approaches depend. CHGI-$\mathbb{R}$ relies on the signal of Sim-RP, while CHGI-$\mathbb{T}$ uses the signal of the holographic TP. When the holographic projection can reproduce the object–image relationship with high fidelity, CHGI-$\mathbb{T}$ has its advantages, such as the sparse matrix case (please see the Exp-RP in Ref.~\cite{Li2025Active}). Conversely, when holographic reconstruction errors are large, as in the case of the intensity-squared chaotic speckle, CHGI-$\mathbb{R}$ proves more reliable. Thus, the relative advantages and disadvantages of the SLM-SPI and DMD-SPI schemes become clearer. When the image quality of the lens imaging system in the DMD-SPI scheme is poor, one can switch to the SLM-SPI scheme instead.

\section{Discussion} 
As is well known, the properties of classical light GI are determined by the projective speckle. For example, the size of speckle grain affects the spatial resolution of the imaging system~\cite{Ferri2010Differential}, and the self-reproduction of speckle pattern enables the copy of ghost images~\cite{Li2019Transverse}, etc. Due to the flexible design of the TP, holographic projection certainly can create many customized reconstruction patterns, which can be further applied to GI scheme.

Figure~\ref{06CopyCase} shows the experimentally measured HBT bunching curves of holographic reconstruction pattern after 10,000 iterations with two types of symmetric mirror TPs. The insets show single-frame positive and negative mirror TPs at the upper right corner of each subgraph in Figs.~\ref{06CopyCase}(a) and \ref{06CopyCase}(b), respectively. Note that, the left half of these TPs originate from the normalized intensity-squared chaotic speckle pattern. Specifically, the positive mirror TP is that the light intensity distribution on the left and right half is completely mirrored. In addition, the negative mirror TP is that the intensity $I(\vec{x})$ of the right half part at $\vec{x}$ is $\beta\times (1-I(-\vec{x}))$, where $\beta =0.2$ is the modulation coefficient to suppress projection intensity and $-\vec{x}$ is the mirror coordinate of $\vec{x}$ in the TP plane. Here, the self-correlated bunching peaks at $x_1=x_2=-1$ mm in Figs.~\ref{06CopyCase}(a) and \ref{06CopyCase}(b) are measured to be 2.41 and 2.77, respectively, which exhibit a super-bunching effect. Meanwhile, the cross-correlated HBT appears at positions $x_1=-x_2=-1$ mm reaching a peak of 1.60 and a valley of 0.92 in Figs.~\ref{06CopyCase}(a) and \ref{06CopyCase}(b), respectively.

Next, we employ the above symmetrical mirror TPs in holographic projection GI. The 0-1 amplitude transmitting letter "N" is placed to the left half of the reconstruction pattern. Here, the TS is the experimentally measured signals which come from the total light intensity transmitting through the detected object "N" at the left half region of the reconstruction pattern. Thus, when the RS is acquired from the left half region, the imaging result is termed an self-correlated ghost image. In contrast, RS comes from the right region, which yields a cross-correlated ghost image. Figure~\ref{07CopyCaseImage} presents three kinds of experimental holographic projection GI (CHGI-$\mathbb{R}$, EHGI-$\mathbb{R}$ and CHGI-$\mathbb{T}$) after 10,000 iterations. According to detected region of TS and RS and the TP category, each holographic projection GI scheme includes four results: positive mirror self-correlated ghost images shown in Figs.~\ref{07CopyCaseImage}(a1)-\ref{07CopyCaseImage}(a3), positive mirror cross-correlated ghost images shown in Figs.~\ref{07CopyCaseImage}(b1)-\ref{07CopyCaseImage}(b3), negative mirror self-correlated ghost images shown in Figs.~\ref{07CopyCaseImage}(c1)-\ref{07CopyCaseImage}(c3), and negative mirror cross-correlated ghost images shown in Figs.~\ref{07CopyCaseImage}(d1)-\ref{07CopyCaseImage}(d3), respectively. To effectively display the imaging results, two color bars are employed for display. As we can see, those experimental results demonstrate the positive and negative copies of ghost image. What's more, even if the low visibility of the valley of cross-correlated HBT bunching effect, the negative mirror cross-correlated ghost image in Fig.\ref{07CopyCaseImage} (d3) by CHGI-$\mathbb{T}$ scheme exhibits noise resistance.

\section{Conclusion} 
In this study, we propose five GI schemes based on holographic projection. Among them, the CHGI-$\mathbb{R}$ scheme and the CHGI-$\mathbb{T}$ scheme are structurally highly similar to two typical SPI schemes, namely SLM-SPI and DMD-SPI. The proposed schemes enable active design of the projection pattern and operate in a lensless configuration. Moreover, holographic projection can be implemented using a high-frame-rate DMD (a liquid crystal phase-only SLM was used due to limitations in the available experimental setup), significantly advancing the potential of practical application of GI schemes our proposed. Experimentally, holographic projection GI based on the intensity-squared speckle and sparse matrix patterns demonstrates a substantial improvement in imaging visibility compared to conventional thermal light GI. Additionally, by artificially customizing the symmetric mirror speckle patterns, we achieved ghost image copy. Notably, the CHGI-$\mathbb{T}$ scheme exhibits a degree of noise resistance.
\begin{acknowledgments}
This work was supported by National Natural Science Foundation of China (Grants No. 62105188, 12175127), Natural Science Foundation of Shandong Province, China (Grant No. ZR2025MS38) and the Scientific Innovation Project for Young Scientists in Shandong Provincial Universities (2024KJG011).
\end{acknowledgments}

\bibliography{refholoprogi.bib}
\end{document}